\documentclass{article}[12pt,a4paper]

\usepackage[fontsize=11pt]{fontsize}
\usepackage[english]{babel}
\usepackage{indentfirst}
\usepackage[letterpaper,top=2cm,bottom=2cm,left=3cm,right=3cm,marginparwidth=1.75cm]{geometry}

\usepackage{amsmath}
\usepackage{bm}
\usepackage{graphicx}
\usepackage[colorlinks=true, allcolors=blue]{hyperref}
\usepackage[numbers,sort & compress]{natbib}
\usepackage{authblk}

\title{ Flow evolution in particle-laden Rayleigh-Bénard convection}
\author{SHI Dai}
\affil[1]{Department of Mechanical Engineering, Osaka University, 2-1 Yamadaoka, Suita, Osaka 565-0871, Japan}
\date{}

\begin{document}
\maketitle

\begin{abstract}

A theoretical analysis employing an averaging ideology is carried out to study flow evolution inside the laminar Rayleigh-Bénard convection system laden with small particles. 
By describing particle dynamics and particle heat as sources of drag and heat respectively, the physics of particle impact on the flow evolution is studied. It is found that due to the relative velocity of the particulate phase to the fluid phase, particles work as a superimposed moment on the whole flow, attenuating the flow intensity.
When the relative temperature of the particulate phase to the fluid phase occurs, particles act like a superimposed moment of buoyancy force on the whole flow, causing alterations in both flow intensity and flow direction.

\end{abstract}

\section{Introduction}
Particle-laden flows with thermally-driven convection involved are important to many fields.
For example, fluids laden with nanoparticles can work as an excellent heat transfer medium with better thermal properties compared to pure fluids, providing opportunities for practical applications involving heat transfer processes, such as the cooling technology in nuclear plants \cite{ref1, ref2, ref3, ref4}.
Another example concerns the particle-based solar collectors where the introduction of small particles into the working fluid of the collector allows improved efficiency and uniformity of the entire heat transfer process, offering possibilities of better utilization of solar energy \cite{ref5, ref6, ref7}. 

The particle-laden flows are one-way, two-way, and four-way coupled, depending on the particle volume fraction \cite{ref8}. 
In the one-way coupled regime, only the particles respond to the flow, not vice versa.
In the two-way coupled regime, not only do the particles respond to the flow, but the flow is also influenced by the particles, resulting in richer phenomena and more physical effects worth studying.
In isothermal two-way coupled systems, flow modifications like turbulence attenuation resulting from the coupled particle dynamics were discovered when the particle size is small \cite{ref9, ref10, ref11, ref12}. The attenuation phenomenon was found to be replaced by a flow enhancement when the particle size became larger, or when the small particles were concentrated into larger clusters \cite{ref9, ref13, ref14}.
When thermally-driven convection is involved, for example, in the two-way coupled Rayleigh-Bénard (RB) convection system, physical effects resulting from the coupled particle heat on the flow motion have also received attention in recent years.
Oresta et al. \cite{ref15, ref16} studied the RB systems laden with particles and found that the heat source contributed by the particles can promote the total Nusselt number (Nu), especially when the particle diameter is small. As particle size increased, this effect caused by the coupled particle heat became less important.
Park et al. \cite{ref17} addressed the importance of the coupled particle heat brought by point particles. 
A significant enhancement of Nu was found to occur with the increase in the particle heat capacity. This enhancement was shown to be pronounced enough that even the attenuation effect caused by the coupled particle dynamics on Nu can be overwhelmed.
Prakhar and Prosperetti \cite{ref18} established a linear theory to study the influence of small particles on the fluid instabilities in the undeveloped RB system. More dimensionless parameters like the particle volume fraction and the normalized particle temperature were found to be critical in describing the two-way coupled thermal system, compared to one-way coupled systems.
Flow modifications caused by finite-sized particles have also been investigated.
Tsutsumi et al. \cite{ref19} found that highly conductive particles can promote the overall heat transfer process inside the RB convection system while particles of low thermal conductivity tended to hinder the heat transfer process by impeding the convective flow. 
Takeuchi et al. \cite{ref20,ref21} studied the laminar RB system laden with finite-sized particles, focusing on the thermal conductivity ratio of particles to the fluid. A typical flow mode of single-direction circulation where the bulk fluid rotated around the domain center along one direction was observed when the particle thermal conductivity was relatively low. 
With an increase in the particle thermal conductivity, the circulation mode was found to transfer to a reversal mode where the flow direction switched with time development. 
The mechanism of the flow evolution, triggered by particles that can reverse the convection direction, is the initial impetus for this study.

The present work focuses on the laminar RB system laden with small particles, and the Euler-Euler viewpoint is adopted to describe this system where the particle number is dense.
Some typical theoretical models to study two-way coupled flows can be reviewed in \cite{ref22, ref23, ref24}. 
For flows laden with small particles, the source flow model, which treats the particles as a point source, was widely adopted and yielded insightful results although sometimes the particles are not really points but finite-sized \cite{ref18, ref25, ref26}. 
In this study, the source flow model is employed to investigate the impacts of coupled particle dynamics and coupled particle heat on flow evolution theoretically.

\section{Theoretical Analysis}
\subsection{Model setup}
\begin{figure}[htbp]
\centering
\includegraphics[width=1\textwidth]{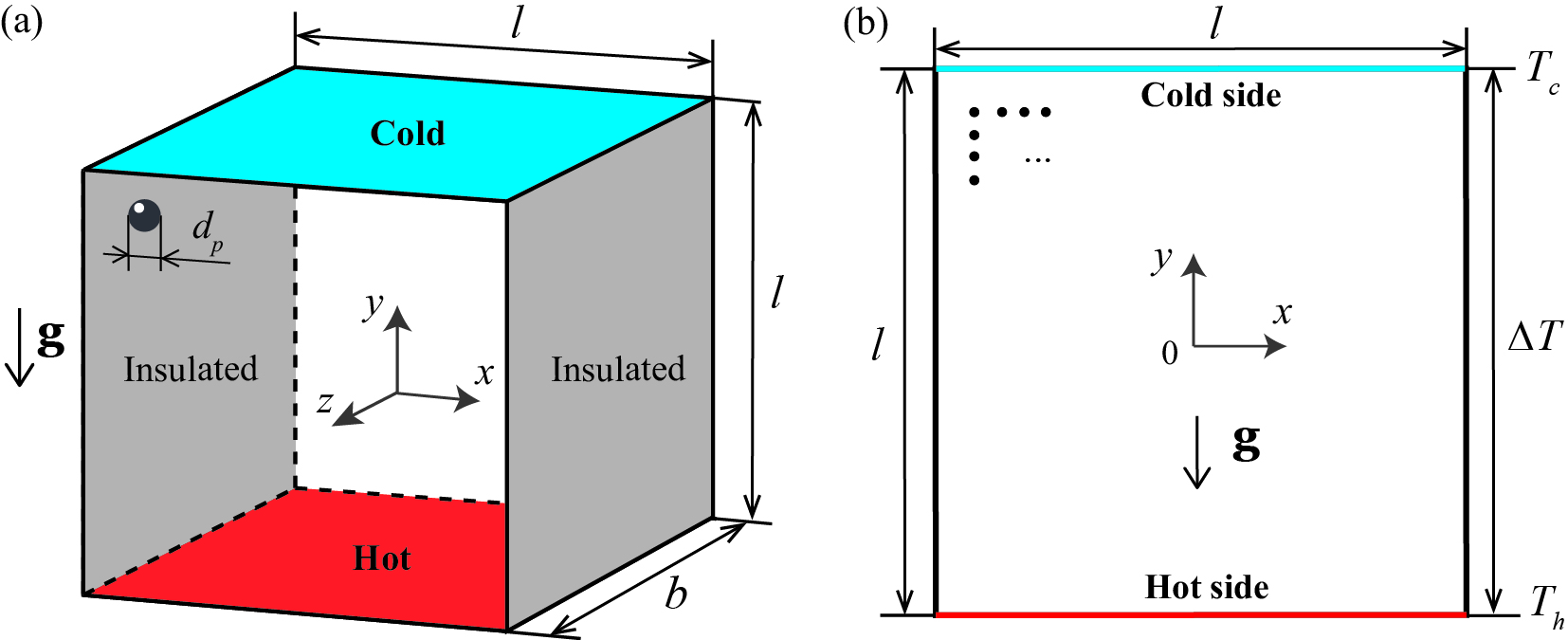}
\caption{\label{F1} The particle-laden RB convection (a) three-dimensional schematic (b) cross-section on the $x \mbox{-} y$ plane }
\end{figure}

An infinitely extended RB convection system along the $z$ direction, where the gravitational acceleration $\bm{g}$ is acting in the $-\bm{e_y}$ direction, is considered in this study, as shown in Fig. \ref{F1}.
The cross-section of the system is a closed square cell of length $l$.
The lateral walls of the system are thermally insulated,
and the temperature difference $\Delta T = T_h - T_c$ between the top and bottom plates is constant.
Mono-dispersed spherical particles of diameter $d_p (\ll l)$ with a dense particle number are suspended in the above RB system.
Throughout this study, the following properties of the fluid and particles are regarded as constant: density $\rho$, kinematic viscosity $\nu$, thermal conductivity $\lambda$, heat diffusivity $\kappa$, volumetric thermal expansion coefficient $\beta$, and specific heat $c$. 
The subscripts $f$ and $p$ indicate the fluid and particle phases, respectively. 

The mass, momentum, and energy equations of the fluid phase are given as follows:
\begin{equation}
    \nabla \cdot \bm{u}=0 ,
\end{equation}
\begin{equation}
    \frac{\partial \bm{u}}{\partial t} + \bm{u} \cdot \nabla \bm{u} = - \frac{1}{\rho_f} \nabla p + \nu_f \nabla^2 \bm{u} + \beta(T_f-T_r) g \bm{e_y} - \eta \frac{\bm{F}}{\rho_f} ,
\end{equation}
\begin{equation}
    \frac{\partial T_f}{\partial t} + \bm{u}\cdot \nabla T_f = \frac{\lambda_f}{\rho_f c_{pf}} \nabla^2 T_f - \eta \frac{ \dot{Q_p}}{\rho_f c_{pf}} ,
\end{equation}
where $\bm{u}$ is the fluid velocity, $p$ is the pressure, $T_f$ is the fluid temperature, \begin{math}
    T_r = \frac{1}{2} \Delta T 
\end{math} is the reference temperature, $\bm{F}$ is the interaction force exerted by the fluid on the particulate phase, and $\dot{Q_p}$ is the heat transfer rate from the fluid to the particulate phase.
In the above, $\eta$ is the factor that describes the volume ratio of the particulate phase to the fluid:
\begin{equation}
    \eta = \frac{V_p}{V} = \frac{v_p N}{V} = v_p n ,
\end{equation}
where $V$ refers to any fluid control volume, $V_p$ represents the corresponding volume occupied by the particulate phase, $v_p$ is the single particle volume, $N$ is the total particle number inside $V$, and $n$ is the particle number density, which varies with time and position.
The particle Reynolds number $ {\rm{Re_p}} = (\epsilon \left\vert u - w \right\vert d_p)/\nu_f $, where $\epsilon$ represents the local void fraction or porosity, is assumed to be small in the laminar system of this study.
Therefore, $\bm{F}$ can be simplified to Stokes drag \cite{ref27}:
\begin{equation}
    \bm{F} = \rho_p \frac{\bm{u-w}}{\tau_p} ,
\end{equation}
where $\bm{w}(\bm{x},t)$ is the velocity of the particulate phase, $\bm{x}$ represents the position vector. The particle relaxation time $\tau_p$ can be written with the general form \cite{ref28}:
\begin{equation}
    \frac{1}{\tau_p} = \frac{3}{4} C_{D} f(\epsilon)^m \frac{\rho_f}{\rho_p} \epsilon^2 \frac{\left |u - w \right |}{d_p},
\end{equation}
where $C_{D}$ is the drag coefficient for single sphere suspension proposed by Schiller and Naumann \cite{ref29},
and $f(\epsilon)^m$ is the porosity function to correct the relaxation time when the particle suspension is not dilute enough. 
The heat transfer rate of $\dot{Q_p}$ can be written as:
\begin{equation}
  \dot{Q_p} = \rho_p c_{pp} \left ( \frac{\partial T_p}{\partial t} + \bm{w} \cdot \nabla T_p   \right ) ,
\end{equation} 
where $T_p(\bm{x},t)$ is the temperature of the particulate phase.
It is noted that $\eta \dot{Q_p}$ can be re-scaled as:
\begin{equation}
    \eta \dot{Q_p} = n m_p c_{pp}  \left ( \frac{\partial T_p}{\partial t} + \bm{w} \cdot \nabla T_p   \right ) = n \pi d_p^2 h_p (T_f - T_p),
\end{equation}
where $m_p$ is the particle mass, and $h_p$ is the particle heat transfer coefficient.
Therefore, the momentum and energy equations of the continuous phase of particles obey the following equations:
\begin{equation}
    \frac{\partial \bm{w}}{\partial t} + \bm{w} \cdot \nabla \bm{w} = \frac{\bm{u-w}}{\tau_p} + (1- \frac{\rho_f}{\rho_p})\bm{g} ,
\end{equation}
\begin{equation}
    \frac{\partial T_p}{\partial t} + \bm{w} \cdot \nabla T_p = \frac{T_f - T_p}{\tau_{th}} ,
\end{equation}
where the particle thermal relaxation time scale $\tau_{th}$ can be given as:
\begin{equation}
    \tau_{th} = \frac{m_p c_{pp}}{\pi d^2_p h_p} = 3 \frac{c_{pp}}{c_{pf}}\frac{{\rm Pr} }{{\rm Nu_p}} \tau_p .
\end{equation}
Here, \begin{math}
  {\rm Pr}= \nu_f /\kappa_f  
\end{math} is the fluid Prandtl number, ${\rm Nu_p}$ is the particle Nusselt number, which comes from the dimensionless process of $h_p$. 
The empirical correlation for ${\rm Nu_p}$ in natural convection systems with immersed spheres can be employed \cite{ref30, ref31}.
Assuming $\rho_p / \rho_f \approx 1 $ and the Stokes drag on the particles are the major driving force, adequate heat exchange between the particulate phase and the fluid can be expected with active phase mixing.
The evolution of $\eta$ can be indicated by the conservation of the total particle number inside the system:
\begin{equation}
    \frac{\partial \eta}{\partial t} + \bm{w} \cdot \nabla \eta + \eta \nabla \cdot \bm{w} = 0.
\end{equation}
Considering the dominant laminar flow in this study, the reference length is set as the cell length of $l$, the characteristic temperature is set as $\Delta T$, the reference velocity is chosen as
\begin{math}
    U= \sqrt{g \beta \Delta T l}
\end{math}, the reference time becomes $l/U$, and the reference pressure is $\rho_f U^2$. Then, the non-dimensional equations for the fluid phase can be given by:
\begin{equation}
    \nabla^* \cdot \bm{u}^*=0 ,
\end{equation}
\begin{equation}
    \frac{\partial \bm{u}^*}{\partial t^*} + \bm{u}^* \cdot \nabla^* \bm{u}^* = -\nabla^* p^* + \sqrt{\frac{{\rm Pr}}{{\rm Ra}}} \nabla^{*2} \bm{u}^* + (T_f^*-T_r)\bm{e_y} + \frac{\rho_p}{\rho_f St_m} \left [  \eta (\bm{w}^* - \bm{u}^*) \right ] ,
\end{equation}
\begin{equation}
    \frac{\partial T_f^*}{\partial t^*} + \bm{u^*}\cdot \nabla^* T_f^* = \frac{1}{\sqrt{{\rm Ra} {\rm Pr}}} \nabla^{*2} T_f^* + \frac{ \rho_p c_{pp} }{\rho_f c_{p f} St_{th}} \left [ \eta (T_p^* - T_f^*) \right] ,
\end{equation}
where the superscript of $*$ indicates the non-dimensional quantities, ${\rm Ra} = \frac{g \beta \Delta T l^3}{\nu_f \kappa_f}$ is the Rayleigh number, $St_m = \frac{\tau_p U}{l}$ is the non-dimensional time ratio of momentum relaxation, and $St_{th} = \frac{\tau_{th} U}{l}$ is the non-dimensional time ratio of thermal relaxation. The last term in the right-hand side of Eq. (14) represents the coupled particle dynamics to the fluid motion while the last term in the right-hand side of Eq. (15) represents the coupled particle heat to the fluid temperature field.
The non-dimensional parameter of the coupled particle heat term can be written as:
\begin{equation}
    \frac{ \rho_p c_{pp} \eta}{\rho_f c_{p f} St_{th}} =  \frac{6 {\rm{Nu_p}}}{\sqrt{{\rm{Ra}}{\rm{Pr}}}} \eta \left( \frac{l}{d_p} \right)^2 .
\end{equation}
Assuming a large number of point-like particles with $\eta \left( \frac{l}{d_p} \right)^2  \gg 1$, the magnitude of the coupled heat term would be much larger than that of the heat conduction term.
Focusing only on first-order terms in the above momentum and energy conservation equations (in other words, performing a linear analysis), 
evolution of the particle-laden flow can be approximately simplified to a group of solvable equations:
\begin{equation}
    \frac{\partial \bm{u}^*}{\partial t^*} = -\nabla^* p^* + \sqrt{\frac{{\rm Pr}}{{\rm Ra}}} \nabla^{*2} \bm{u}^* + (T_f^*-T_r)\bm{e_y} + \frac{\rho_p}{\rho_f St_m} \left [  \eta (\bm{w}^* - \bm{u}^*) \right ] ,
\end{equation}
\begin{equation}
    \frac{\partial T_f^*}{\partial t^*} = \frac{ \rho_p c_{pp} }{\rho_f c_{p f} St_{th}} \left [ \eta (T_p^* - T_f^*) \right] ,
\end{equation}
\begin{equation}
    \frac{\partial \bm{w^*}}{\partial t^*} = \frac{\bm{u^*-w^*}}{St_m} ,
\end{equation}
\begin{equation}
    \frac{\partial T_p^*}{\partial t^*} = \frac{T_f^* - T_p^*}{St_{th}} ,
\end{equation}
\begin{equation}
     \frac{\partial \eta}{\partial t^*} + \bm{w^*} \cdot \nabla^* \eta + \eta \nabla^* \cdot \bm{w^*} = 0 .
\end{equation}

\subsection{Averaging process}
\begin{figure}[htbp]
\centering
\includegraphics[width=1\textwidth]{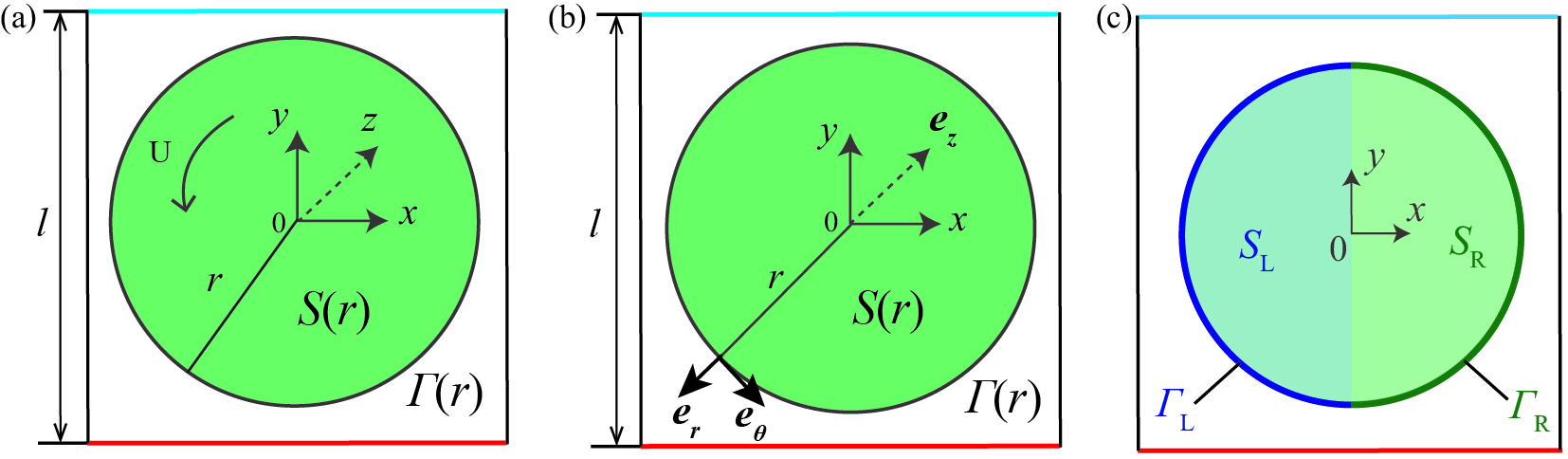}
\caption{\label{F2} Convection region on the $x \mbox{-} y$ plane (a) the control volume of $S$ (b) the oriented surface (c) the symmetrical parts of $S$ and $\Gamma$}
\end{figure}

A coherent convective flow along the angular direction, whose reference velocity is $U$, is assumed to rotate around the domain center of the closed container in the cross-section and infinitely extends in the $z$ direction. For simplicity, we only focus on the cross-section of $x \mbox{-} y$ plane, as shown in Fig. \ref{F2}(a).
A circular control volume of $S(r)$ (the green part), where $r$ is the radial coordinate, is supposed to cover the convective flow. A space-averaging process over $S$, which is bounded by a closed curve of $\Gamma(r)$, is carried out in this part.

Based on $S$, an oriented surface with the surface element vector of $\bm{e}_z d S$ can be defined. This oriented surface is enclosed by an oriented curve, whose unit vector is $\bm{e}_{\theta}$, as shown in Fig. \ref{F2}(b).
Taking the curl of Eq. (17) and integrating the result over this oriented surface gives:
\begin{equation}
    \frac{\partial}{\partial t} \langle u_{\theta} \rangle_\Gamma = \sqrt{ \frac{{\rm{Pr}}} {{\rm{Ra}}}  } \left ( \frac{\partial^2}{\partial r^2} \langle u_{\theta} \rangle_\Gamma + \frac{1}{r}\frac{\partial}{\partial r}\langle u_{\theta} \rangle_\Gamma  \right ) +
    \frac{1}{\beta g} \langle F_{b \theta} \rangle_{\Gamma} + \frac{\rho_p}{\rho_f St_m} \left [    \langle \eta w_{\theta} \rangle_{\Gamma}   -   \langle \eta u_{\theta} \rangle_{\Gamma}  \right ]   , 
\end{equation}
where $\langle \cdot \cdot \cdot \rangle _ {\Gamma}$ denotes the curve average. In the above, $\langle u_\theta \rangle_\Gamma (r,t)$ is the curve-averaged velocity of the fluid phase in the angular direction, which comes from:
\begin{equation}
    \int_s (\nabla \times \bm{u}) \cdot \bm{e_z} dS = \int_\Gamma \bm{u} \cdot \bm{e_\theta} d \Gamma = \int_\Gamma u_\theta d \Gamma = \Gamma \langle u_\theta \rangle_\Gamma .
\end{equation}
A positive $\left \langle u_{\theta} \right \rangle_\Gamma$ suggests the flow rotates along $\Gamma$ in the counterclockwise direction while a negative $\left \langle u_{\theta} \right \rangle_\Gamma$ suggests the flow rotates in the clockwise direction. 
Evolution of the curve-averaged velocity $\langle w_\theta \rangle_\Gamma (r,t)$  of the particulate phase  can be given after a similar treatment of Eq. (19):
\begin{equation}
    \frac{\partial}{\partial t} \langle w_\theta \rangle_\Gamma = \frac{\langle u_\theta \rangle_\Gamma - \langle w_\theta \rangle_\Gamma }{St_m} .
\end{equation}
The first term on the right-hand side of Eq. (22) comes from the viscous term:
\begin{equation}
\begin{split}
  \int_s \nabla \times (\nabla^2\bm{u}) \cdot \bm{e_z} dS 
  &= \int_{\Gamma} (\nabla^2\bm{u}) \cdot \bm{e_\theta} d \Gamma
  \\ 
  &= \int_0^{2 \pi} r \left ( 
  \frac{\partial^2 \bm{u}}{\partial r^2} \cdot \bm{e}_\theta + \frac{1}{r}\frac{\partial \bm{u}}{\partial r} \cdot \bm{e}_\theta + \frac{1}{r^2}\frac{\partial^2 \bm{u}}{\partial \theta^2} \cdot \bm{e}_\theta
  \right ) d \theta
   \\
   &= \Gamma \left (  \frac{\partial^2}{\partial r^2} \langle u_\theta \rangle_\Gamma + \frac{1}{r} \frac{\partial}{\partial r} \langle u_\theta \rangle_\Gamma   \right ).
\end{split}    
\end{equation}
The second term of $\langle F_{b \theta} \rangle_\Gamma (r,t)$ on the right-hand side of Eq. (22) comes from:
\begin{equation}
     \int_s \left [ \nabla \times (T_f-T_r) \bm{e}_y \right ] \cdot \bm{e_z} dS =
     \frac{1}{\beta g} \int_s (\nabla \times \bm{F}_{b} ) \cdot \bm{e_z} dS =  \frac{1}{\beta g} \int_\Gamma (\bm{F}_{b} \cdot \bm{e_\theta}) d \Gamma =  \frac{\Gamma}{\beta g}  \langle F_{b \theta} \rangle_\Gamma ,
\end{equation}
where $\bm{F}_{b} = \beta g (T_f - T_r) \bm{e_y}$ represents the buoyancy force. 
It is found that $\langle F_{b \theta} \rangle_\Gamma$ can also be related to the surface-averaged horizontal temperature gradient of the fluid phase:
\begin{equation}
    \frac{\Gamma}{\beta g}  \langle F_{b \Gamma} \rangle_\Gamma =  \int_s [ \nabla \times (T_f \bm{e_y}) ] \cdot \bm{e_z} dS =  \int_s  \frac{\partial T_f}{\partial x}\bm{e_z}  \cdot \bm{e_z} dS = S \left \langle \frac{\partial T_f}{\partial x} \right \rangle _S .
\end{equation}
Therefore, if the fluid energy equation (18) is differentiated with respect to $x$ and integrated over the entire control volume of $S$, we find:
\begin{equation}
    \frac{d}{d t} \left \langle \frac{\partial T_f}{\partial x} \right \rangle _S = \frac{\rho_p c_{pp}}{\rho_f c_{p f} St_{th}} \left \langle \frac{\partial \left [ \eta (T_p - T_f)\right ]}{\partial x} \right \rangle _S  ,
\end{equation}
where $\langle \cdot \cdot \cdot \rangle _ {S}$ denotes the surface average. 
With a similar treatment of Eq. (19), the surface-averaged horizontal temperature gradient of the particulate phase can be given as:
\begin{equation}
    \frac{d}{d t} \left \langle \frac{\partial T_p}{\partial x} \right \rangle_S = \frac{1}{St_{th}} 
    \left ( \left \langle \frac{\partial T_f}{\partial x} \right \rangle_S -  \left \langle \frac{\partial T_p}{\partial x} \right \rangle_S \right ).
\end{equation}

\section{Results and Discussion}
Through the averaging process, the flow evolution problem can be further simplified to the solution of five differential equations with five undetermined variables. 
However, instead of describing results under various initial conditions, we want to focus more on the physical meaning of the particulate phase on flow evolution.
\subsection{Physics of coupled particle dynamics}
Flow evolution with the involvement of coupled particle dynamics can be summarized from Eq. (22) and Eq. (24), where the first term in Eq. (22) represents the time evolution of the flow velocity, and the last term refers to the coupled particle dynamics via the Stokes drag.
This result becomes easier to read if we start by neglecting the impact of particles:
\begin{equation}
    \frac{\partial}{\partial t} \langle{u_{\theta}}\rangle  = \sqrt{ \frac{{\rm{Pr}}} {{\rm{Ra}}}  } \left ( \frac{\partial^2}{\partial r^2} \langle u_{\theta} \rangle_\Gamma + \frac{1}{r}\frac{\partial}{\partial r}\langle u_{\theta} \rangle_\Gamma  \right )  + \frac{1}{\beta g} \langle{F_{b \Gamma}}\rangle.
\end{equation}
Eq. (30) reminds us of the single-phase momentum equation, where the buoyancy force drives and affects the convective flow inside the RB system. 
The scaling of buoyancy force to viscous force is $\sqrt{{Ra}/{Pr}}$, indicating the convection would develop more strongly with the increase of $Ra$ or the decrease of $Pr$.
Now, let us turn to the converse case, where only the particle term is considered:
\begin{equation}
    \frac{\partial}{\partial t} \langle u_{\theta} \rangle_\Gamma = \frac{\rho_p}{\rho_f St_m} \left [    \langle \eta w_{\theta} \rangle_{\Gamma}   -   \langle \eta u_{\theta} \rangle_{\Gamma}  \right ]   . 
\end{equation}
It is shown that the particle dynamics acts like a superimposed particulate momentum caused by the relative particle velocity on the fluid momentum, working to modify the flow velocity. When particle distribution is regarded as almost uniform, Eq. (31) can be further simplified as:
\begin{equation}
    \frac{\partial}{\partial t} \langle{u_{\theta}}\rangle_\Gamma  =  \frac{\rho_p (1-\epsilon )}{\rho_f St_m} \left [\langle{w_{\theta}\rangle_{\Gamma} }  - \langle{u_{\theta}}\rangle_{\Gamma} \right ]  ,
\end{equation}
where $1-\epsilon$ refers to the solid volume fraction.
Given a quiescent base state, the bulk fluid is invoked from rest to form the coherent convection by the imbalance in the buoyancy force, and $\langle{u_{\theta}}\rangle - \langle{w_{\theta}}\rangle > 0 $ is suggested during the time development, leading to a decreasing trend of $\langle u_{\theta} \rangle$. 
Therefore, the coupled dynamics of point-like particles act like a hindering factor to the convection intensity, barely altering the convection direction.

\subsection{Physics of coupled particle heat}
The coupled particle heat affects the flow intensity and direction via the buoyancy term in Eq. (22), the process of which obeys Eqs. (27)$\mbox{-}$(29). 
This role played by the coupled particle heat can be further understood by looking at its physical picture:
\begin{equation}
   \frac{\rho_p c_{pp}}{\rho_f c_{p f} St_{th}}  \frac{\partial \left [ \eta (T_p - T_f)\right ]}{\partial x}  = \frac{\rho_p c_{pp}}{\rho_f c_{p f} St_{th}} \left [ \frac{\partial \eta}{\partial x} (T_p - T_f) + \eta \frac{\partial (T_p - T_f)} {\partial x} \right ].
\end{equation}

\begin{figure}[htbp]
\centering
\includegraphics[width=0.5\textwidth]{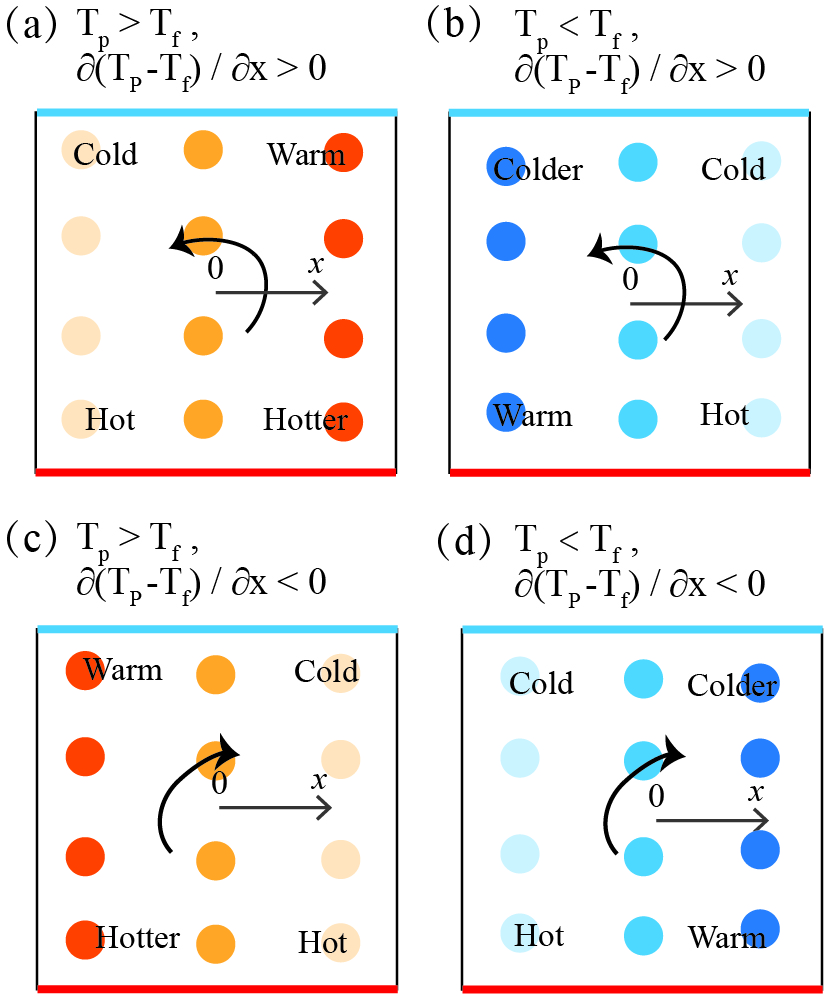}
\caption{\label{F3} The physical meaning of $A$ }
\end{figure}

\begin{figure}[htbp]
\centering
\includegraphics[width=0.5\textwidth]{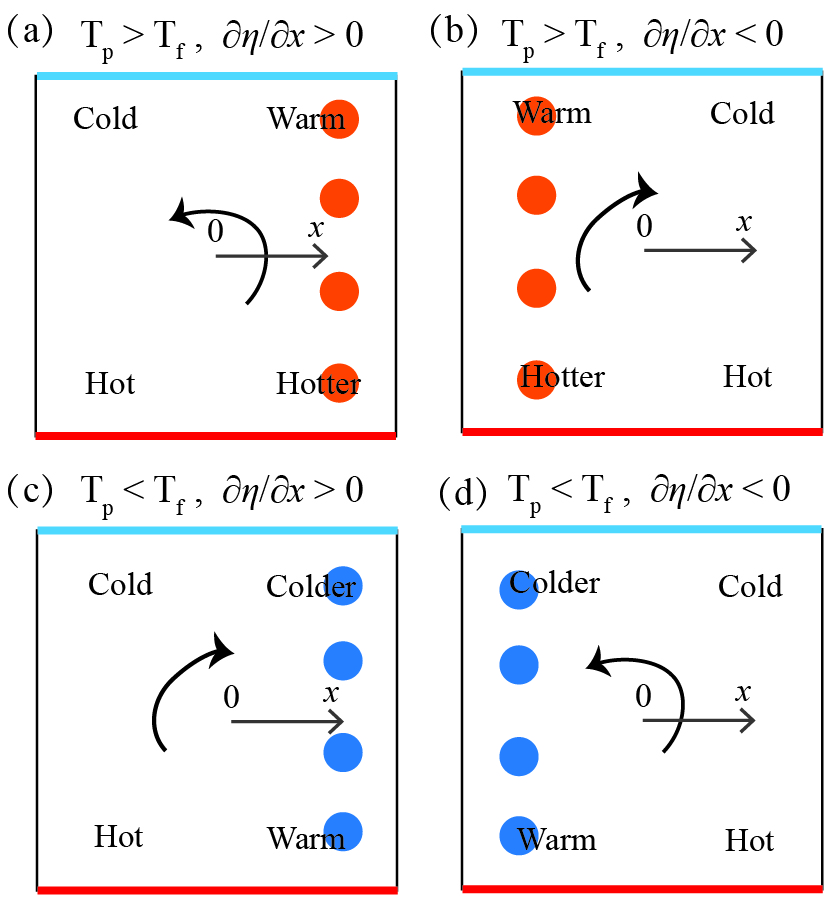}
\caption{\label{F4} The physical meaning of $B$ }
\end{figure}

In the last term of Eq. (33), $\eta$ is considered independent of $x$. As a result, a uniform particle number distribution in the horizontal direction can be assumed, with the focus switched to $A = \frac{\partial (T_p - T_f)} {\partial x}$.
Under the circumstance of $A > 0, T_p - T_f > 0$, relatively hot particles can be imagined as introduced into the system from the right side wall, resulting in a superimposed thermal plume that tends to rotate in the counterclockwise direction on the current flow, as shown in Fig. \ref{F3}(a). As a result, the flow is to be altered at the next time point until this relative particulate temperature relaxes to zero.
Under the circumstance of $A > 0, T_p - T_f < 0$, relatively cold particles can be imagined as introduced into the system from the left side wall, superimposing a counterclockwise thermal plume to the current flow, as shown in Fig. \ref{F3}(b). Similar cases of $ A < 0 $ are explained in Fig. \ref{F3}(c)$\mbox{-}$(d).
Now, let us look at the other term of $ B= \frac{\partial \eta}{\partial x}(T_p - T_f) $ in Eq. (31), where $T_p - T_f$ is considered independent of $x$. 
As a result, $T_p - T_f> 0$ can be imagined as $T_p > T_h$ and $T_p - T_f< 0$ can be imagined as $T_p < T_c$.
Therefore, under the circumstance of $B>0, T_p - T_f > 0$, more hot particles are supposed to locate on the right side of the system, generating an ascending thermal plume in the right half of the system, as shown in Fig. \ref{F4}(a).
Under the circumstance of $B<0, T_p - T_f > 0$, these hot particles are transported to the left side of the system, generating an ascending thermal plume in the left half of the system, as shown in Fig. \ref{F4}(b).
Similar analysis of $T_p - T_f < 0$ are explained in  Fig. \ref{F4}(c)$\mbox{-}$(d).
Therefore, the coupled particle heat works to modify the flow motion with time development until the relative particle temperature is relaxed to zero. 

The influence of couple particle heat can be further imagined by connecting the evolution of the moment of buoyancy force.
Curve-averaged moment of buoyancy force $\langle M_f \rangle_\Gamma $ of the fluid phase can be written as:
\begin{equation}
  \langle M_f \rangle_\Gamma = \frac{1}{\Gamma} \int_\Gamma \beta g (T_f-T_r) x d \xi,
\end{equation}
where $\xi$ is the integration variable referred to $\Gamma$.
Treating $\Gamma$ as the sum of two symmetrical parts $\Gamma_L$ and  $\Gamma_R$, as shown in Fig. \ref{F2}(c), $\langle M_f \rangle_\Gamma$ can be rewritten as:
\begin{equation}
  \langle M_f \rangle_\Gamma 
  = \frac{\beta g}{\Gamma} 
  \left [ \int_{\Gamma_L}  T_f x d\xi + \int_{\Gamma_R}  T_f x d\xi \right ] 
  = \frac{\beta g}{2} \langle x \rangle_{\Gamma R} \ [ \langle T_f \rangle_{\Gamma R} -  \langle T_f \rangle_{\Gamma L} ] .
\end{equation}
Strictly speaking, there should be a factor of $1- \eta$ on the right-hand of Eq. (32), as the integral operation actually refers to the whole region of $\Gamma$, not the region occupied by the fluid phase within $\Gamma$.
However, it is assumed that the fluid phase occupies the majority of $\Gamma$ so Eq. (32) is still valid.
Similarly, the curve-averaged moment of buoyancy force contributed by the particulate phase can be written as:
\begin{equation}
      \langle M_p \rangle_\Gamma 
      = \frac{1}{\Gamma} \int_\Gamma \eta \beta g (T_p-T_f - T_r) x d \xi   \\
      = \frac{\beta g }{2}  \langle x \rangle_{\Gamma R} \ \left [
      \langle \eta (T_p - T_f) \rangle_{\Gamma_R}-\langle \eta (T_p - T_f) \rangle_{\Gamma_L}
      \right ].
\end{equation}
Treating the control volume of $S$ as the sum of two symmetrical halves of $S_L$ and $S_R$, as shown in Fig. \ref{F2}(c), the fluid energy equation (28) can be rewritten in the curve-averaged form as:
\begin{equation}
    \frac{d}{d t} \left [  \langle T_f \rangle_{\Gamma_R} - \langle T_f \rangle_{\Gamma_L} \right]
    = \frac{\rho_p c_{pp}}{\rho_f c_{p f} St_{th}} \left [  \langle \eta (T_p - T_f) \rangle_{\Gamma_R} - \langle \eta (T_p - T_f) \rangle_{\Gamma_L}     \right ] . 
\end{equation}
Using Eqs. (35)$\mbox{-}$(37), we have:
\begin{equation}
    \frac{d}{d t} \langle M_f \rangle_\Gamma = \frac{\rho_p c_{pp}}{\rho_f c_{p f} St_{th}} \langle M_p \rangle_\Gamma .
\end{equation}
Eq. (38) clearly exhibits the function played by the coupled particle heat, which works to superimpose a moment of buoyancy force $\langle M_p \rangle_\Gamma$ on the flow, altering both magnitude and direction of the convective flow at the next time point.
When the sign of $\langle M_p \rangle_\Gamma$ and $\langle M_f \rangle_\Gamma$ are the same, the convective flow tends to be further enhanced by the particulate phase. When the sign of $\langle M_p \rangle_\Gamma$ and $\langle M_f \rangle_\Gamma$ are opposite, the convective flow tends to be attenuated by the particles, resulting in phenomena like flow reversal in extreme cases.

\section{Conclusion}
This study establishes a mathematical model of the particle-laden RB convection with an Euler-Euler viewpoint. An averaging process over the convection region is employed to investigate flow evolution under both impacts of coupled particle dynamics and coupled particle thermal.
For small particles, $w_{\theta}$ follows well with the flow velocity $u_{\theta}$. As a result, the coupled particle dynamics works as a superimposed particulate momentum that directly affects the flow velocity, causing a flow attenuation phenomenon while barely altering the flow direction. 
The coupled particle heat works as a superimposed moment of buoyancy force that indirectly affects the flow velocity via the buoyancy term, altering both flow intensity and flow direction.
With the increase of solid volume fraction and the density ratio of particle to fluid, or the decrease of $St_m$, the influence of coupled particle dynamics is assumed to be more obvious.
With the increase of heat capacity ratio of particle to fluid, or the decrease of $St_{th}$, the influence of coupled particle heat is assumed to be more pronounced. 

These results highlight the impact of the particulate phase on the flow performance, which contributes to our physical understanding of the two-way coupled effects in convection systems.

\section*{Acknowledgments}
This work was supported by JST SPRING, Grant Number JPMJSP2138.


\begin{thebibliography}{31}
\bibitem{ref1} P. Keblinski, J. A. Eastman, and D. G. Cahill. Nanofluids for thermal transport. Material Today, 8(2005), 36–44.
\bibitem{ref2} A. A. A. Arani, M. Mahmoodi, and S. M. Sebdani. On the Cooling Process of Nanofluid in a square enclosure with linear temperature distribution on left wall. Journal of Applied Fluid Mechanics, 7(2014), 591-601.
\bibitem{ref3} Z. Haddad, H. F. Oztop, A review on natural convective heat transfer of nanofluids. Renewable Sustainable Energy Reviews. 16(2012), 5363-5378.
\bibitem{ref4} J. Ahuja and J. Sharma., Rayleigh-Bénard instability in nanofluids: a comprehensive review. Micro and Nano Systems Letters, 8:21(2020), 1-15.
\bibitem{ref5} H. Pouransari et al., Effects of preferential concentration on heat transfer in particle-based solar receivers. Journal of Solar Energy Engineering, 139(2017), 1-11.
\bibitem{ref6} K. H. Clifford, Advances in central receivers for concentrating solar applications. Solar Energy, 152(2017), 38-56.
\bibitem{ref7} K. H. Clifford et al., Review of high-temperature central receiver designs for concentrating solar power. Renewable Sustainable Energy Reviews, 29(2014), 835–846.
\bibitem{ref8} S. Elghobashi, On predicting particle-laden turbulent flows. Applied Scientific Research, 52(1994), 309-329.
\bibitem{ref9} R. A. Gore et al., Effect of particle size on modulating turbulent intensity. International Journal of Multiphase Flow, 15(2)(1989), 279-285.
\bibitem{ref10} J. D. Kulick et al., Particle response and turbulence modification in fully developed channel flow. Journal of Fluid Mechanics, 277(1994), 109-134.
\bibitem{ref11} S. V. Apte et al., Large-eddy simulation of swirling particle-laden flows in a coaxial-jet combuster. International Journal of Multiphase Flow, 29(2003), 1311-1331.
\bibitem{ref12} W. Hwang et al., Homogeneous and isotropic turbulence modulation by small heavy (St$\sim$50) particles. Journal of Fluid Mechanics, 564 (2006), 361–393.
\bibitem{ref13} T. Bosse et al., Small particles in homogeneous turbulence: Settling velocity enhancement by two-way coupling. Physics of Fluids, 18(027102)(2006), 1-17.
\bibitem{ref14} A. Aliseda et al., Effect of preferential concentration on the settling velocity of heavy particles in homogeneous isotropic turbulence. Journal of Fluid Mechanics, 468 (2002), 77–105.
\bibitem{ref15} P. Oresta et al., Heat transfer mechanisms in bubbly Rayleigh-Bénard convection. Physical Review E, 80(026304)(2009), 1-11.
\bibitem{ref16} P. Oresta et al., Effects of particle settling on Rayleigh-Bénard convection. Physical Review E, 87(063014)(2013), 1-11.
\bibitem{ref17} H. J. Park et al., Rayleigh-Bénard turbulence modified by two-way coupled inertial, nonisothermal particles. Physical Review Fluids, 3(034307)(2018), 1-15.
\bibitem{ref18} S. Prakhar et al., Linear theory of particulate Rayleigh-Bénard instability. Physical Review Fluids, 6(083901)(2021), 1-19.
\bibitem{ref19} T. Tsutsumi et al., Heat transfer and particle behaviors in dispersed two-phase flow with different heat conductivities for liquid and solid. Flow Turbulence Combust, 92(2014), 103–119.
\bibitem{ref20} S. Takeuchi et al., Effect of temperature gradient within a solid particle on the rotation and oscillation modes in solid-dispersed two-phase flows. International Journal of Heat and Fluid Flow, 43(2013), 15-25.
\bibitem{ref21} S. Takeuchi et al., Flow reversals in particle-dispersed natural convection in a two-dimensional enclosed square domain. Physical Review Fluids, 4(084304)(2019), 1-22.
\bibitem{ref22} D. A. Drew, Mathematical modeling of two-phase flow. Annual Review of Fluid Mechanics, 15(1983), 261-291.
\bibitem{ref23} D. Migdal et al., A source flow model for continuum gas-particle flow. Journal of Applied Mechanics, 12(1967),860-865.
\bibitem{ref24} P. Oresta et al., Multiphase Rayleigh-Bénard convection. Mechanical Engineering Reviews, 1(2014), 1-18.
\bibitem{ref25} M. Boivin et al., On the prediction of gas-solid flows with two-way coupling using large eddy simulation. Physics of Fluids, 12(2000),2080-2090.
\bibitem{ref26} S. Balachandar, A scaling analysis for point-particle approaches to turbulent multiphase flows. International Journal of Multiphase Flow, 35(2009), 801-810.
\bibitem{ref27} M. R. Maxey et al., Equation of motion for a small rigid sphere in a nonuniform flow. The Physics of Fluids, 26(1983), 883-889.
\bibitem{ref28} L. Mazzei and P. Lettieri. A drag force closure for uniformly dispersed fluidized suspensions. Chemical Engineering Science, 62(2007), 6129-6142.
\bibitem{ref29} L. Schiller et al., A drag coefficient correlation. Zeitschrift des Vereins Deutscher Ingenieure, 77(1933), 318-320. 
\bibitem{ref30} S. W. Churchill, Free convection around immersed bodies. Heat Exchanger Design Handbook, Begell House, New York, 2002, Section 2.5.7.
\bibitem{ref31} F. P. Incropera et al., Fundamentals of Heat and Mass Transfer, John Wiley and Sons, New Jersey, 2007, 583-583.


\end{thebibliography}
\end{document}